\documentclass[12pt]{article}
\usepackage{amsmath}
\usepackage{environ}
\usepackage{etoolbox}
\usepackage{hyperref}
\hypersetup{colorlinks,linkcolor={blue},citecolor={blue},urlcolor={blue}}
\usepackage[pdftex]{graphicx}
\usepackage{graphics}
\usepackage[version=4]{mhchem}
\usepackage{pict2e}
\usepackage{mathtools, nccmath}
\usepackage{diffcoeff}
\usepackage{setspace}
\usepackage{etoc}
\usepackage{cite}
\usepackage{float}
\usepackage{lipsum}
\usepackage{blindtext}
\usepackage[table,xcdraw]{xcolor}
\usepackage{multirow}
\usepackage{booktabs}
\usepackage{siunitx}
\usepackage{booktabs}
\usepackage{outline}
\usepackage{avant} 
\usepackage{multirow}
 \usepackage[table,xcdraw]{xcolor}

\usepackage{array}
\usepackage{rotating}
\usepackage{tikz}
\usepackage{lscape}
\usepackage{subcaption}
\usepackage[toc,page]{appendix}
\usepackage{enumitem}

\title{\textbf{Muon ($g-2$) in $U(1)_{L_{\mu}-L_{\tau}}$ Scotogenic Model Extended with Vector like Fermion}}

\author {Simran Arora\thanks{ 009simranarora@gmail.com}, Monal Kashav\thanks{ monalkashav@gmail.com}, Surender Verma\thanks{s\_7verma@hpcu.ac.in} and B. C. Chauhan\thanks{ bcawake@hpcu.ac.in}}

\date{\textit{Department of Physics and Astronomical Science,\\Central University of Himachal Pradesh, Dharamshala 176215, INDIA.}}

\begin{document}
\maketitle

\begin{abstract}
 The latest results of anomalous muon magnetic moment at Fermilab show a discrepancy of 4.2 $\sigma$ between the Standard Model (SM) prediction and experimental value. In this work, we revisit $U(1)_{L_{\mu}-L_{\tau}}$ symmetry with in the paradigm of scotogenic model which explains muon ($g-2$) and neutrino mass generation, simultaneously. The mass of new gauge boson $M_{Z_{\mu\tau}}$ generated after the spontaneous symmetry breaking of $U(1)_{L_{\mu}-L_{\tau}}$ is constrained, solely, in light of the current neutrino oscillation data to explain muon ($g-2$). In particular, we have obtained two regions I and II, around 150 MeV and 500 MeV, respectively, in $M_{Z_{\mu\tau}}-g_{\mu\tau}$ plane which explain the neutrino phenomenology. Region I is found to be consistent with muon neutrino trident (MNT) bound ($g_{\mu\tau}$ $\leq$ $10^{-3}$) to explain muon ($g-2$), however, region II violates it for mass range $M_{Z_{\mu\tau}}>300$ MeV. We, then, extend the minimal gauged scotogenic model by a vector like lepton (VLL) triplet $\psi_T$. The  mixing of $\psi_T$ with inert scalar doublet $\eta$ leads to chirally enhanced positive contribution to muon anomalous magnetic moment independent of $Z_{\mu\tau}$ mass. Furthermore, we have, also, investigated the implication of the model for $0\nu\beta\beta$ decay and $CP$ violation. The non-observation of $0\nu\beta\beta$ decay down to the sensitivity of 0.01 eV shall refute the model. The model, in general,is found to be consistent with both $CP$ conserving and $CP$ violating solutions.   
   
\end{abstract}
\section{Introduction}

The standard model (SM) of particle physics has been very successful in explaining the observed  dynamics of fundamental particles and their interactions. The discovery of Higgs boson at large hadron collider (LHC) has affirmed our belief in the theory though there are few questions which still remain unanswered as of now. For example, observation of non-zero neutrino mass, matter-antimatter asymmetry, existence of dark matter (DM), to name a few, are some of the astounding  unresolved issues in the SM and require new physics scenarios for their explanation. Recently, combining the previous result of Brookhaven National Laboratory (BNL) \cite{Muong-2:2006rrc}, the muon $(g-2)$ collaboration at Fermilab has reported a 4.2$\sigma$ discrepancy between the experimental observation ($a_\mu^{exp}$) and SM prediction ($a_\mu^{th}$) of muon anomalous magnetic moment $a_\mu$. More precisely, they found $\Delta a_{\mu}=a_\mu^{exp}-a_\mu^{th}=(2.51\pm 0.59)\times 10^{-9}$\cite{Muong-2:2021ojo} which is a sign of new physics motivating theoretical initiatives to calculate various contributions to muon ($g-2$) with extremal precision, in particular, the quantum chromodynamics (QCD) contribution\cite{Borsanyi:2020mff}. 

Several new physics scenarios have been proposed which can cause muon ($g-2$) to differ from the SM prediction. In supersymmetric models, particles like smuon, neutralino, chargino enter in the loop and can give additional contribution to muon ($g-2$) \cite{Martin:2001st}. Also, there are plethora of models which attempt to explain DM and contributions to muon ($g-2$). In some of these models DM particle participate in the loop diagram and contribute to muon ($g-2$). There are another class of models where DM particle do not directly interact with the SM particles, alternatively, mediator particle interacts with both DM and SM particle(s). The mediator particle can give contribution to muon ($g-2$) \cite{Pospelov:2008zw}. Furthermore, leptoquark models can, also, give contribution to muon ($g-2$) through the diagrams involving leptoquarks and SM quarks \cite{Biggio:2014ela}. The two Higgs doublet model (2HDM) and its variations have, also, been proposed for the explanation of muon ($g-2$) \cite{Han:2015yys,Chen:2021jok,Arcadi:2021yyr,Dey:2021pyn,DelleRose:2020oaa,Iguro:2019sly,Chun:2020uzw}.

Apart from the above possibilities, another class of model extends the SM with anomaly free $U(1)_{L_{\mu}-L_{\tau}}$ \cite{He:1990pn,He:1991qd} symmetry for the explanation of muon ($g-2$)\cite{Borah:2021mri,Biswas:2016yan, Zhou:2021vnf,Guo:2006qa,Dev:2017fdz, Majumdar:2020xws,Patra:2016shz}. After spontaneous breaking of $U(1)_{L_{\mu}-L_{\tau}}$ symmetry, the new boson $Z_{\mu\tau}$ participate in the loop diagram to give additional contribution to muon ($g-2$). Further variation of $U(1)_{L_{\mu}-L_{\tau}}$ model is the gauged extension with in the framework of scotogenic model. The gauged scotogenic model is promising variation which explain neutrino mass generation, dark matter and muon ($g-2$), simultaneously.  These scenarios have been extensively explored in different variations e.g. see Refs. \cite{Borah:2021khc,Jana:2020joi,Baek:2015fea,Han:2019diw,Kang:2021jmi}. The neutrino masses are generated by usual scotogenic process at one loop level by requiring three right handed neutrinos and an inert doublet. It is known that, in this framework, the mass of $Z_{\mu\tau}$ ($M_{Z_{\mu\tau}}$) has to be less than the bound $\mathcal{O}(10^2)$ MeV coming from muon neutrino trident (MNT) process\cite{Altmannshofer:2014pba}. Although there are severe experimental constraints on new gauge boson mass from CCFR\cite{Altmannshofer:2014pba}, WD cooling\cite{Bauer:2018onh,Kamada:2018zxi}, COHERENT\cite{COHERENT:2017ipa,COHERENT:2020iec}, NA62 \cite{Krnjaic:2019rsv} and NA64 \cite{Gninenko:2014pea} but there still remain allowed parameter space in $M_{Z_{\mu\tau}}-g_{\mu\tau}$ plane which can explain muon ($g-2$). There are few attempts to widen the mass range of $Z_{\mu\tau}$ explaining muon ($g-2$) while still being consistent with MNT and other experimental bounds\cite{Cheng:2021okr, Chen:2021jok}. In general, the explanation of muon $g-2$ based on gauged scotogenic model assumes small mixing between $Z_{\mu\tau}$ and SM gauge boson $Z$. Otherwise the new gauge boson may not explain the muon ($g-2$) due to constraints from Z-pole precision observable at LEP\cite{Freitas:2014pua}. 

In the present work, we revisit the possible explanation of muon ($g-2$) in $U(1)_{L_\mu-L_\tau}$ scotogenic extension of the SM model wherein the particle content has been enlarged with three right handed neutrino ($N_k$, $k=e,\mu,\tau$), one scalar singlet $S$ and one inert doublet $\eta$. $N_k$ and $\eta$ are odd under unbroken $Z_2$ symmetry making them suitable DM matter candidate while scalar singlet $S$ is even. Unlike the earlier works, the mass range of new gauge boson $M_{Z_{\mu\tau}}$ explaining muon ($g-2$) has been constrained in light of the current neutrino oscillation data. Also, as discussed earlier, the explanation of muon $g-2$ based on gauged scotogenic model assumes small mixing between $Z_{\mu\tau}$ and SM gauge boson $Z$. In this work, we implement an extension of the gauged scotogenic model with vector like lepton (VLL) triplet $\psi_T$ wherein muon ($g-2$) can be explained independent of above mixing pattern and $U(1)_{L_\mu-L_\tau}$ gauge boson mass. In fact, the mixing of $\psi_T$ and inert doublet $\eta$ of scotogenic model results in chirally enhanced positive contribution to muon ($g-2$).



The paper is organised as follows. In Section 2, we discuss minimal gauged scotogenic model. In Section 3, we extend the minimal gauged scotogenic model, discussed in Section 2, by a vector like lepton triplet and show that it successfully explains muon ($g-2$) and neutrino oscillation data. Finally, in Section 4, we brief our conclusions.

\section{Minimal Gauged Scotogenic Model}
  Scotogenic model proposed by E. Ma\cite{Ma:2006km} is a promising framework to explain  dark matter (DM) and non-zero neutrino mass, simultaneously. Within this model, the field content is enlarged with three right handed neutrinos and an inert doublet $\eta$. In general, the contribution of inert doublet $\eta$ to muon ($g-2$) is negative\cite{Jana:2020joi,Queiroz:2014zfa,Calibbi:2018rzv}. This motivates to study $U(1)_{L_{\mu}-L_{\tau}}$ extensions of scotogenic model for possible explanation of muon ($g-2$) and neutrino phenomenology.  The SM gauge group is extended by $U(1)_{L_{\mu}-L_{\tau}}$ symmetry.  The complete particle content of the model with corresponding charge assignments are given in Table \ref{tab1}. Here,  $ L_{k}(k=e,\mu,\tau)$ are the usual SM left handed lepton doublets, $k_{R}(k=e,\mu,\tau)$ are SM right handed charged leptons, $N_{k}(k=e,\mu,\tau)$ are right handed neutrino singlets, $H$ is SM Higgs, $\eta$ is inert doublet and $S$ is scalar singlet field introduced to break $U(1)_{L_{\mu}-L_{\tau}}$ symmetry to induce the mass of the new gauge boson $Z_{\mu\tau}$. All the beyond SM (BSM) fields are odd under $Z_{2}$ except $S$, the lightest of which can be DM candidate within the model.

\begin{table}[t]
\centering
 \begin{tabular}{l c c c c c c} 
 \hline

 Symmetry Group & $L_{e}$, $L_{\mu }$, $L_{\tau }$ & $e_{R}$, $\mu_{R}$, $\tau_{R}$ &$N_{e}$, $N_{\mu }$, $N_{\tau }$ & $H$ & $S$ & $\eta$ \\ [0.5ex] 
 \hline\hline
 $SU(2)_{L}$ $\times$ $U(1)_{Y}$ & (2, -1/2) & (1, -1) & (1, 0) &(2, 1/2) & (1, 1)&(2, 1/2) \\ 
$ U(1)_{L\mu-L\tau}$ &(0, 1, -1) & (0, 1, -1) & (0, 1, -1) &0&1&0 \\
$Z_{2}$ & + & + & -& + & + & - \\ [1ex] 
 \hline
 \end{tabular}
 \caption{The field content and respective charge assignments under $ SU(2)_{L}\times U(1)_{Y}\times U(1)_{L_{\mu}-L_{\tau}}\times Z_{2}$.}
      \label{tab1}
\end{table}

The relevant terms in Lagrangian can be written as 
\begin{equation}\label{lagfull}
    \mathcal{L} =   \mathcal{L}_{scalar} +\mathcal{L}_{N} - \frac{1}{4} (Z_{\mu \tau})_{\mu\nu}Z^{\mu \nu}_{\mu \tau} - \frac{\epsilon}{2} (Z_{\mu \tau})_{\mu\nu}B^{\mu\nu},
    \end{equation}
    where $\mathcal{L}_{Scalar}$ is the Lagrangian containing kinetic  and potential terms of the scalar sector ($H,\eta,S$) of the model, $\mathcal{L}_{N}$ is the Lagrangian for neutrino sector. The fourth and fifth terms in Eqn. (\ref{lagfull}) are, kinetic term for $Z_{\mu \tau}$ and mixing term of $Z_{\mu \tau}$-$Z$ gauge bosons, respectively ($\epsilon$ is mixing parameter).
    The Lagrangian for scalar fields is given by
\begin{eqnarray}
\nonumber
\mathcal{L}_{Scalar} &=& (\mathcal{D}_{\mu}H)^{\dagger}(\mathcal{D}_{\mu}H)+(\mathcal{D}_{\mu}\eta)^{\dagger}(\mathcal{D}_{\mu}\eta) + (\mathcal{D}_{\mu}S)^{\dagger}(\mathcal{D}_{\mu}S) \nonumber\\ && -\: V(H,\eta,S),
 \end{eqnarray}

  where the covariant derivative $\mathcal{D}_{\mu}$ is given by

\begin{eqnarray}
 \mathcal{D}_{\mu}& =& \partial_{\mu} -i\frac{g}{2}\tau .W_{\mu} - i g^{'} \frac{Y}{2}B_{\mu} - i g_{\mu\tau}Y_{\mu\tau}(Z_{\mu \tau})_{\mu}
\end{eqnarray}
and the scalar potential $V(H,\eta,S)$ can be written as

  \begin{eqnarray}
    \nonumber
    V(H,\eta, S) & = & -\mu_{H}^{2}(H^{\dagger} H) + \mu_{\eta}^{2}(\eta^{\dagger}\eta)- \mu_{S}^{2}(S^{\dagger}S) + \lambda_{1}(H^{\dagger} H)^{2}   + \lambda_{2}(\eta^{\dagger}\eta)^{2} 
    \nonumber \nonumber \\ && 
+\: \lambda_{3}(\eta^{\dagger}\eta)(H^{\dagger} H)  +\lambda_{4}(\eta^{\dagger} H)(H^{\dagger} \eta)+\frac{\lambda_{5}}{2}[(H^{\dagger} \eta)^{2}
 \\
     && \: +(\eta^\dagger H)^{2}]+ \lambda_{S}(S^{\dagger
}S)^{2} + \lambda_{HS}(H^{\dagger} H)(S^{\dagger} S) +
     \lambda_{\eta S}(\eta^{\dagger} \eta)(S^{\dagger} S).
\end{eqnarray}
The SM gauge symmetry is broken by the neutral component of Higgs doublet $H$ while the $ U(1)_{L_{\mu}-L_{\tau}}$ symmetry is broken by the non-zero vacuum expectation value ($vev$) $v_{S}$ of scalar singlet $S$  which results in a massive gauge boson $Z_{\mu\tau}$ with mass $M_{Z_{\mu\tau}} = g_{\mu\tau}v_{s}$ where $g_{\mu\tau}$ is $L_{\mu}-L_{\tau}$ gauge coupling. Also, we assume $\mu_{\eta}^2$ $>$ 0  so that $\eta$ do not acquire any $vev$. After the symmetry breaking, neutral components of the scalar fields can be written as

\begin{align}
  \nonumber
   H^{0} &= \frac{1}{\sqrt{2}}(v+h_{1}+i h_{2}), \\ \nonumber
\eta^{0} &=  \frac{1}{\sqrt{2}}(\eta_{1} + i \eta_{2}), \\
       S &= \frac{1}{\sqrt{2}}(v_{S}+s_{1} + i s_{2}), 
\end{align}

 where $\left\langle {S}\right\rangle = v_{S}/\sqrt{2}$, $\left\langle {H}\right\rangle = (0,v/\sqrt{2})^{T}$  with $v=246$ GeV. \\
In the basis \{$ h_{1}$, $s_{1}$, $\eta_{1}$, $\eta_{2}$\}, mass matrix for the neutral scalars is
 
 \begin{align}
M_{0}^{2}&= \begin{pmatrix}
M_{11} & M_{12} & 0 & 0\\
M_{12} & M_{22} & 0 & 0 \\
0 & 0& M_{33} &  M_{34} \\
0 & 0 & M_{34} & M_{44}\\
\end{pmatrix},
\end{align}
where

\begin{align}
\nonumber
    M_{11} &= 2\lambda_{1}v^2 , \\
    \nonumber
    M_{12} &= \lambda_{HS}vv_{S} ,\\
    \nonumber
    M_{22} &= 2\lambda_{S}v_{S}^{2},\\
    \nonumber
     M_{33} &= \mu_{\eta}^2 + \frac{1}{2}\left( \lambda_{3}+\lambda_{4}+ Re[\lambda_{5}]\right)v^2 + \frac{1}{2}\lambda_{\eta S} v_{S}^2,\\
     \nonumber
         M_{34} &= -Im[\lambda_{5}]v^{2},\\
         \nonumber
      M_{44} &=\mu_{\eta}^2 + \frac{1}{2}( \lambda_{3}+\lambda_{4}- Re[\lambda_{5}])v^2 + \frac{1}{2}\lambda_{\eta S}v_{S}^2 .
\end{align}
The mass matrix for the neutral scalar can be diagonalized by 
\begin{equation}
   \begin{pmatrix}
h  \\s   \\

\end{pmatrix} =  \begin{pmatrix}
\cos\alpha_{1} & -\sin\alpha_{1} \\
\sin\alpha_{1} & \cos\alpha_{1} \\

\end{pmatrix}
\begin{pmatrix}
h_{1} \\
s_{1}  \\

\end{pmatrix},
\end{equation}
and 
\begin{equation}
   \begin{pmatrix}
\eta_{R}^{0}  \\
\eta_{I}^{0}   \\

\end{pmatrix} =  \begin{pmatrix}
\cos\alpha_{2} & -\sin\alpha_{2}\\
\sin\alpha_{2} & \cos\alpha_{2} \\

\end{pmatrix}
\begin{pmatrix}
\eta_{1} \\
\eta_{2} \\

\end{pmatrix},
\end{equation}
where $\tan \alpha_{1,2} =\frac{r_{1,2}}{1+\sqrt{1+r_{1,2}^{2}}}$  with $ r_{1} = \frac{vv_{S}\lambda_{HS}}{\lambda_{1}v^{2}-\lambda_{S}v_{S}^{2}}$ and $r_{2}=-\frac{Im[\lambda_{5}]}{Re[\lambda_{5}]}$ \cite{Cabral-Rosetti:2017mai}. Hence masses of scalars are

\begin{equation}
  M_{s,h}^{2} = \lambda_{1}v^{2} + \lambda_{S}v_{S}^2 \pm (\lambda_{1}v^{2} + \lambda_{S}v_{S}^{2})\sqrt{1+r_{1}^{2}} ,
\end{equation}
 
 while the masses of charged scalar $H^{\pm}$, pseudoscalars $\eta_{I}^{0}$ and $\eta_{R}^{0}$ are given by 
 \begin{eqnarray}
   & M_{H^{\pm}}^{2} &= \mu_{\eta}^{2} + \frac{1}{2}(\lambda_{3}v^{2} + \lambda_{\eta S}v_{S}^{2}),\\
\nonumber
   & M_{\eta_{I}^{0},\eta_{R}^{0}}^{2} &= M_{H^{\pm}}^{2} + \left(\frac{\lambda_{4}}{2} \pm |\lambda_{5}|\right)v^{2},
 \end{eqnarray}
 where $+$($-$) sign is for $\eta_{I}^{0}$($\eta_{R}^{0}$).

Furthermore, the Yukawa Lagrangian for the model is given by 

    \begin{eqnarray} \label{nlag}
\nonumber
-\mathcal{L}_{N} &\supset & y_{e} \bar{L}_{e} H e_{R} +y_{\mu} \bar{L}_{\mu} H \mu_{R} + y_{\tau} \bar{L}_{\tau} H \tau_{R} \nonumber \\  && 
+\: y_{\eta e} \bar{L}_{e} \tilde{\eta} N_{e} + y_{\eta \mu} \bar{L}_{\mu} \tilde{\eta} N_{\mu } + y_{\eta \tau} \bar{L}_{\tau} \tilde{\eta} N_{\tau}  \nonumber \\ && +\: y_{e\tau} S  N_{e} N_{\tau} + y_{e\mu} S^{*} N_{e} N_{\mu}   \nonumber \\ &&+\:  \frac{M_{ee}}{2} N_{e} N_{e} + M_{\mu\tau} N_{\mu} N_{\tau}+ h.c ,
\end{eqnarray}

where $\tilde{\eta}=i \sigma_{2}\eta^{*}$. Using Eqn. (\ref{nlag}), the charged lepton mass matrix $M_{l}$,  Dirac Yukawa matrix $y_{D}$ and right handed neutrino mass matrix $M_{R}$ are given by

\begin{eqnarray*}
M_{l} = \frac{1}{\sqrt{2}}\begin{pmatrix}
   y_{e}v && 0 && 0 \\
   0 &&  y_{\mu}v && 0 \\
   0 && 0 &&  y_{\tau}v
   \end{pmatrix},
\end{eqnarray*}

\begin{eqnarray}
  y_{D} = 
   \begin{pmatrix}
   y_{\eta e} && 0 && 0 \\
   0 &&  y_{\eta \mu} && 0 \\
   0 && 0 &&  y_{\eta \tau}
   \end{pmatrix}  , M_{R} =
    \begin{pmatrix}
   M_{ee} && x && y \\
   x &&  0 && M_{\mu\tau}e^{i\delta} \\
   y && M_{\mu\tau}e^{i\delta} && 0
   \end{pmatrix},
\end{eqnarray}
where $x\equiv y_{e\mu}v_{S}/\sqrt{2}$, $y\equiv y_{e\tau}v_{S}/\sqrt{2}$ and $\delta$ is the phase remaining after redefinition of the fields.

\subsection{Neutrino Masses and muon ($g-2$)}
\begin{figure}[t]
    \centering
    \includegraphics[width=12cm]{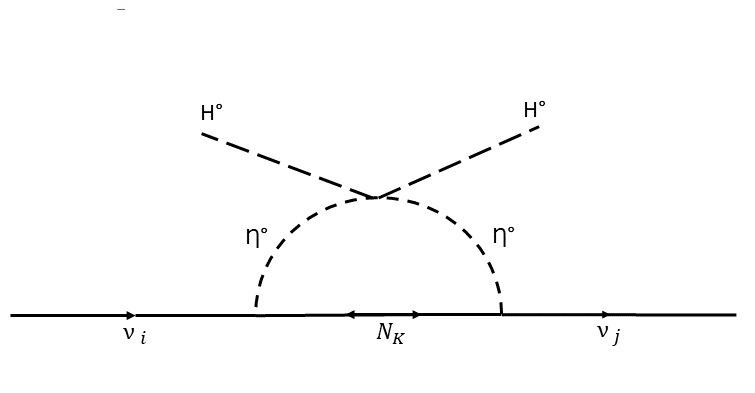}
    \caption{The diagram responsible for neutrino mass generation in scotogenic model at one loop level. }
    \label{fig:1}
\end{figure}

The neutrino masses are generated at one loop level (Fig. \ref{fig:1}) resulting in the light neutrino mass matrix given by\cite{Ma:2006km, Merle:2015ica}
 
 \begin{equation}\label{scot}
     M_{ij}^{\nu} = \sum_{k}{}\frac{y_{ik}y_{jk}M_{k}}{16\pi^{2}}\left[\frac{M_{\eta_{R}^{0}}^{2}}{M_{\eta_{R}^{0}}^{2}-M_{k}^{2}}\ln\frac{M_{\eta_{R}^{0}}^{2}}{M_{k}^{2}}-\frac{M_{\eta_{I}^{0}}^{2}}{M_{\eta_{I}^{0}}^{2}-M_{k}^{2}}\ln\frac{M_{\eta_{I}^{0}}^{2}}{M_{k}^{2}}\right],
 \end{equation}
  where $M_{k}$ is the mass of $k^{th}$ right handed neutrino and $M_{\eta_{R}^{0},\eta_{I}^{0}}$ are the masses of real and imaginary parts of inert doublet $\eta$. The Yukawa couplings appearing in Eqn. (\ref{scot}) are derived from $y_D$ in the basis where $M_{R}$ is diagonal. 
 If the mass squared difference between $\eta_{R}^{0}$ and $\eta_{I}^{0}$  i.e. ${M_{\eta_{R}^{0}}^{2}}-M_{\eta_{I}^{0}}^{2} = \lambda_{5}v^{2}$  $<<$  $M^{2}$ where $M^2 = (M_{\eta_{R}^{0}}^{2}-M_{\eta_{I}^{0}}^{2})/2$, then above expression reduces to 
 \begin{equation}\label{mmatrix}
    M_{ij}^{\nu}=\frac{\lambda_{5} v^{2}}{16 \pi^{2}}\sum_{k}{}\frac{y_{ik}y_{jk}M_{k}}{M^{2}-M_{k}^{2}}\left[1-\frac{M_{k}^{2}}{M^{2}-M_{k}^{2}}\ln\frac{M^{2}}{M_{k}^{2}}\right].
\end{equation}

The effective low energy neutrino mass matrix $M^\nu$ obtained using Eqn. (\ref{mmatrix}) can be diagonalized to ascertain model predictions for neutrino masses and mixing angles $viz.,$
\begin{equation}
    M^\nu=U M^\nu_d U^T,
\end{equation}
where $U$ is unitary matrix and $M^\nu_d=diag(m_1,m_2,m_3)$, $m_i$ are neutrino mass eigenvalues.  
In term of the elements of the diagonalizing matrix
\begin{eqnarray}
   U =
    \begin{pmatrix}
   U_{e1} && U_{e2} && U_{e3} \\
     U_{\mu 1} && U_{\mu 2} && U_{\mu 3} \\
    U_{\tau 1} && U_{\tau 2} && U_{\tau 3} 
   \end{pmatrix},
\end{eqnarray}
the neutrino mixing angles can be evaluated using

\begin{eqnarray}
    \sin^{2}\theta_{13} = |U_{e3}|^{2} ,\quad   \sin^{2}\theta_{23} = \frac{|U_{\mu 3}|^{2}}{1 - |U_{e3}|^{2}} ,\quad \sin^{2}\theta_{12} = \frac{|U_{e2}|^{2}}{1 - |U_{e3}|^{2}}.
\end{eqnarray}

\begin{figure}
    \centering
    \includegraphics[width=8cm]{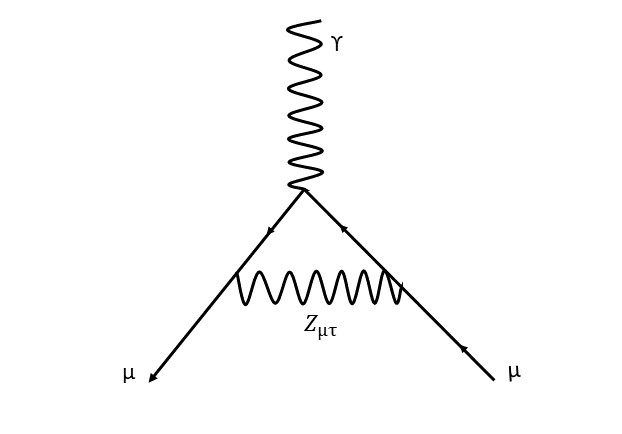}
    \caption{One loop contribution to muon ($g-2$) mediated by neutral gauge boson $Z_{\mu\tau}$. }
    \label{fig:3}
\end{figure}

\textbf{$Z_{\mu\tau}$ contribution to Muon ($g-2$):}

 It is to be noted that though the charged component of inert doublet at one loop level contributes negatively to the muon magnetic moment but is too small and there is dominant positive contribution from $Z_{\mu\tau}$ (Fig. \ref{fig:3}). The one loop contribution to muon anomalous magnetic moment $\Delta a_{\mu}$ arising from  $U(1)_{L_{\mu}-L_{\tau}}$ neutral gauge boson $Z_{\mu\tau}$ is given by\cite{Baek:2008nz,Brodsky:1967sr} 

\begin{equation}\label{au}
    \Delta a_{\mu}(Z_{\mu\tau}) = \frac{g_{\mu\tau}^2}{8\pi^{2}} \int_{0}^{1} dz \frac{2 m_{\mu}^{2} z^{2} (1-z)}{ z^{2} m_{\mu}^{2}  + (1-z) M_{Z_{\mu \tau}}^{2} }  \approx \frac{g_{\mu \tau}^{2}}{8 \pi^{2}} \frac{2 m_{\mu}^{2}}{3 M_{Z_{\mu \tau}}^{2}},
\end{equation}
where $M_{Z_{\mu\tau}}$ is the mass of gauge boson $(Z_{\mu\tau})$ and $m_\mu$ is mass of muon with $M_{Z_{\mu\tau}}>m_\mu$. 
The coupling strength $g_{\mu\tau}$ of gauge boson is severely constrained to be less than $10^{-3}$ from the measurement of muon neutrino trident (MNT) cross-section by experiments like CCFR\cite{Altmannshofer:2014pba}, COHERENT \cite{COHERENT:2017ipa,COHERENT:2020iec}, BABAR \cite{BaBar:2016sci}.

  \begin{table}[t]
   
      \centering
        \begin{tabular}{ll}
             \hline\hline
       Parameter  &  Range \\
       \hline
       
        $M_{ee}$ & [30, 70] GeV \\

         $M_{\mu\tau}$ & [30, 80] GeV \\

          $x$ &  [30, 80] GeV \\ 
           $y$ & [20, 60] GeV \\ 
         $ \delta$ & [0, 2$\pi$] \\ 
             $y_{\eta k}$  & $[10^{-3}, 10^{-1}]$  \\
                $\mu_{\eta}$ & $[10, 10^3]$ GeV \\ 
                $ \lambda_{3}$ & $[10^{-2}, 10^{-1}]$ \\ 
                  $\lambda_{4}$ & $[10^{-1}, 1]$ \\ 
                   $ \lambda_{5}$ & $[10^{-5}, 10^{-4}]$\\ 
                 $\lambda_{\eta S}$ & $[10^{-2}, 10^{-1}]$ \\ 
                  $v_{S}$ & $[10^{2}, 9\times 10^{2}]$ GeV \\
                 
                     $ N_{k}$ &$ [10^{6}, 10^{8}]$ GeV  \\ 
                     
                       \hline
        \end{tabular}
   
    \caption{The parameter ranges used in the numerical analysis ($k=e,\mu,\tau$).}
    \label{tab2}
\end{table}

\subsection{Numerical Analysis} \label{sec2.2}
The mass of the gauge boson $M_{Z_{\mu\tau}}=g_{\mu\tau}v_S$ where the $vev$ $v_S$, also, appears in $M_R$. Consequently, when we write Dirac Yukawa matrix $y_D$ in $M_R$ diagonal basis, the low energy effective neutrino mass matrix $M^\nu$ (Eqn. (\ref{mmatrix})) depends on $v_S$ through couplings $y_{ik}$ and $y_{jk}$. The range of $v_S$ is constrained by demanding the model to have consistent low energy phenomenology. The neutrino masses and mixing angles are obtained by diagonalising the neutrino mass matrix. In the numerical analysis, the model predictions for neutrino mass squared differences ($\Delta m^{2}_{21}=m_2^2-m_1^2,|\Delta m^{2}_{31}|=|m_3^2-m_1^2|$) and mixing angles ($\theta_{13},\theta_{23},\theta_{12}$) are compared with the experimental data ($3\sigma$ ranges)\cite{Esteban:2020cvm} 
 \begin{equation*}
    \sin^{2} \theta_{13} = (0.02034-0.02430) ,\quad  \sin^{2} \theta_{23} = (0.407 - 0.620) ,\quad \sin^{2} \theta_{12} = (0.269 - 0.343) ,
    \end{equation*}
    \begin{equation}\label{ndata}
  \Delta m^{2}_{31} = (2.431 - 2.599) \times 10^{-3} eV^{2} , \quad \Delta m^{2}_{21} = (6.82 - 8.04) \times 10^{-5} eV^{2},
    \end{equation}
    to constrain the allowed parameter space.
    The free parameters in the neutrino mass matrix are varied randomly with in ranges given in Table \ref{tab2}.

\begin{figure}[t]
    \centering
 \includegraphics[width=8.5cm]{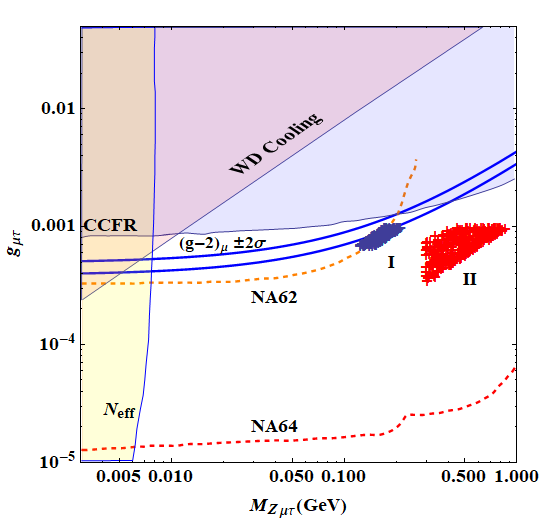}
    \caption{The allowed parameter space consistent with low energy neutrino phenomenology in $M_{Z_{\mu\tau}}-g_{\mu\tau}$ plane. The parameter space in region I around 150 MeV (blue) is consistent with the experimental observation of muon ($g-2$)(solid blue lines).}
    \label{fig3}
\end{figure}

The allowed parameter space, in $M_{Z_{\mu\tau}}$-$g_{\mu\tau}$ plane, which satisfies the low energy neutrino oscillation data and its prediction for muon ($g-2$) is shown in Fig \ref{fig3}. It is evident from Fig. \ref{fig3} that there exist two regions in $g_{\mu\tau}-M_{Z_{\mu\tau}}$ plane, $M_{Z_{\mu\tau}}\approx$ 150 MeV and 500 MeV, for which neutrino phenomenology is satisfied within the model. The experimental bounds from CCFR\cite{Altmannshofer:2014pba}, WD cooling \cite{Bauer:2018onh,Kamada:2018zxi} and future sensitivities of experiments like NA62 \cite{Krnjaic:2019rsv} and NA64 \cite{Gninenko:2014pea} are, also, shown (dashed lines) in Fig {\ref{fig3}}. The upper triangular region is excluded from cooling of white dwarf systems (WD) \cite{Bauer:2018onh,Kamada:2018zxi}. It can be seen from Fig. \ref{fig3} that region I (around 150 MeV) explains the muon ($g-2$) while region II satisfies neutrino phenomenology but not muon ($g-2$). Hence, the minimal $U(1)_{L_{\mu}-L_{\tau}}$ scotogenic model explains muon ($g-2$) for $M_{Z_{\mu\tau}}\approx$ 150 MeV. In the next section, we propose a framework in which we can explain muon ($g-2$) in region II i.e. in absence of contribution from $Z_{\mu\tau}$.

\textbf{Benchmark Point:} In order to emphasise the viability of the model based on correct low energy neutrino phenomenology and experimental measurement of muon ($g-2$), we have obtain the benchmark point of the numerical analysis for representative values of input parameters given in Table \ref{tab3}. The model predictions for neutrino mixing angles and mass squared differences are 

 \begin{equation*}
    \sin^{2} \theta_{13} = 0.020 ,\quad  \sin^{2} \theta_{23} = 0.56,\quad \sin^{2} \theta_{12} = 0.30 ,
    \end{equation*}
    \begin{equation*}
  \Delta m^{2}_{31} = 2.4\times 10^{-3} eV^{2} , \quad \Delta m^{2}_{21} = 7.8\times 10^{-5} eV^{2}.
    \end{equation*}
The value of $\Delta a_{\mu} (Z_{\mu\tau}) =2.78\times 10^{-9} $ for $M_{Z_{\mu\tau}} = 147$ MeV and $g_{\mu\tau} = 0.0008$.

 \begin{table}[t]
   
      \centering
        \begin{tabular}{ll}
             \hline
             \hline
       Parameter  &  Value \\
       \hline
       
        ($M_{ee}$,  $M_{\mu\tau}$)  & (69.2, 66) GeV \\


          ($x, y$) &  (54.8, 26.6) GeV \\ 
         $ \delta$ & 347.52$^o$ \\ 
             $y_{\eta k}$  & (0.043, 0.048, 0.047)  \\
                ($M_{\eta_{R}^{0}}, M_{\eta_{I}^{0}}$) & (172.97, 172.98) GeV \\ 
                  $v_{S}$ & 184 GeV \\
                     $ N_{k}$  & (7.96, 2.04, 6.77) $\times 10^{7}$ GeV  \\ 
                       \hline
        \end{tabular}
   
    \caption{The values of parameters used to obtain benchmark point of the numerical analysis ($k=e,\mu,\tau$).}
    \label{tab3}
\end{table}

\section{Extension of Minimal Gauged Scotogenic Model by a Vector like Lepton (VLL) triplet}\label{sec3}

In general, the explanation of muon $g-2$ based on gauged scotogenic model\cite{Borah:2021khc,Jana:2020joi,Baek:2015fea} assumes small mixing between $Z_{\mu\tau}$ and SM gauge boson $Z$. In absence of the $Z_{\mu\tau}$ contribution (large $M_{Z_{\mu\tau}}$ or large mixing paradigms), we extend the gauged scotogenic model by a VLL triplet $\psi_{T}$ to explain muon ($g-2$)
\begin{eqnarray}
\psi_{T} = 
   \begin{pmatrix}
   \frac{\psi_{T}^{^{-}}}{\sqrt{2}} && \psi_{T}^{^{0}}  \\
  \psi_{T}^{^{--}} &&   -\frac{\psi_{T}^{^{-}}}{\sqrt{2}} \\
   \end{pmatrix},
  \end{eqnarray}

with charge assignments $(3,-1,1)$ under $SU(2)_L\times U(1)_Y\times U(1)_{L_\mu-L_\tau}$ symmetry and odd under $Z_{2}$.

The $L_{\mu}-L_{\tau}$ charge of $\psi_{T}$ allows it couples to muon only and shall not contribute in neutrino mass generation. The new terms in the Lagrangian (Eqn. (\ref{lagfull})) are 
 \begin{eqnarray} \label{lpsi}
\mathcal{L}_{\psi_{T}} = \bar{\psi_{T}}i\gamma^{\mu}\mathcal{D_{\mu}}\psi_{T} - y_{\psi} \eta^{\dagger}\bar {\psi}_{T,R}L_{\mu} - M_{\psi} \bar{\psi_{T}} \psi_{T} + h.c.
 \end{eqnarray}
 where $M_{\psi} \bar{\psi_{T}} \psi_{T}$ is the bare mass term for $\psi_{T}$.
 
\begin{figure} [t]
    \centering
    {{\includegraphics[width=5cm]{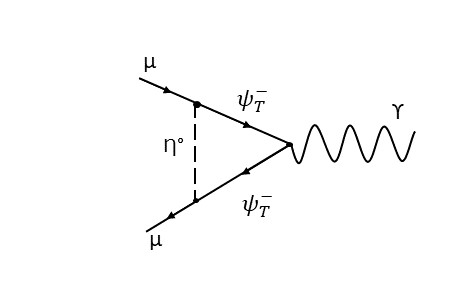} }}%
    \qquad
    {{\includegraphics[width=5cm]{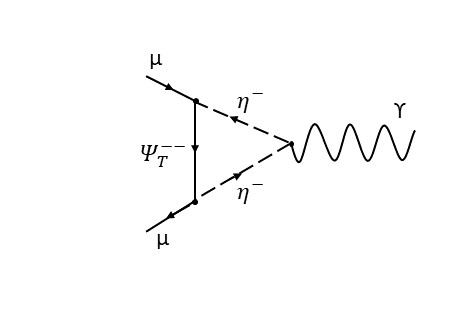}}}
    \qquad
    {{\includegraphics[width=5cm]{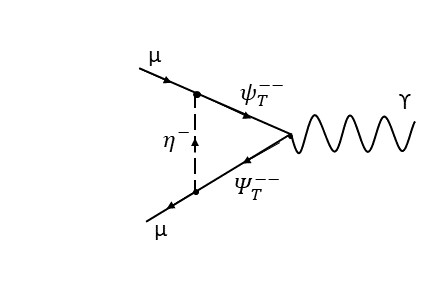}}}
    \caption{The diagrams responsible for positive contribution, from $\psi_{T}$ and $\eta$, to muon ($g-2$) at one loop level.}%
    \label{fig44}%
\end{figure}

The $SU(2)_L$ triplet $\psi_{T}$ alone gives negative contribution to muon ($g-2$)\cite{Freitas:2014pua}. Also, as discussed in the previous section, the charged component of scalar doublet $\eta$ has a negative contribution to muon ($g-2$). However, $\psi_{T}$ coupling with $\eta$, may results in chirally enhanced positive contribution to $\Delta a_{\mu}$ through the second term of Eqn. (\ref{lpsi}). The possible diagrams contributing to muon ($g-2$) are shown in Fig. \ref{fig44}. 

The contribution to muon magnetic moment is given by \cite{Freitas:2014pua}:
\begin{eqnarray} \label{au2}
 \Delta a_{\mu} (\eta + \psi_{T}) = \frac{m_{\mu}^2 y_{\psi}^2}{16 \pi^2 M_{\eta}^2}[5 F_{FFS}(M_{\psi}^2/M_{\eta}^2) - 2 F_{SSF}(M_{\psi}^2/M_{\eta}^2)],
\end{eqnarray}
where 
$M_{\psi}$, $M_{\eta}$ are the masses of VLL triplet $\psi_{T}$ and inert scalar doublet $\eta$, $y_\psi$ is coupling constant, 
\begin{eqnarray*}
  F_{FFS}(t) = \frac{1}{6(t-1)^4}[t^3 - 6t^2 + 3t + 2 + 6t \ln t], 
\end{eqnarray*}
and
\begin{eqnarray}
   F_{SSF}(t) = \frac{1}{6(t-1)^4}[-2t^3 - 3t^2 + 6t - 1 + 6t^2 \ln t],
\end{eqnarray}
where $t=\frac{M_{\psi}^2}{M_{\eta}^2}$. 

\begin{figure} [t]
    \centering
    {{\includegraphics[width=7cm]{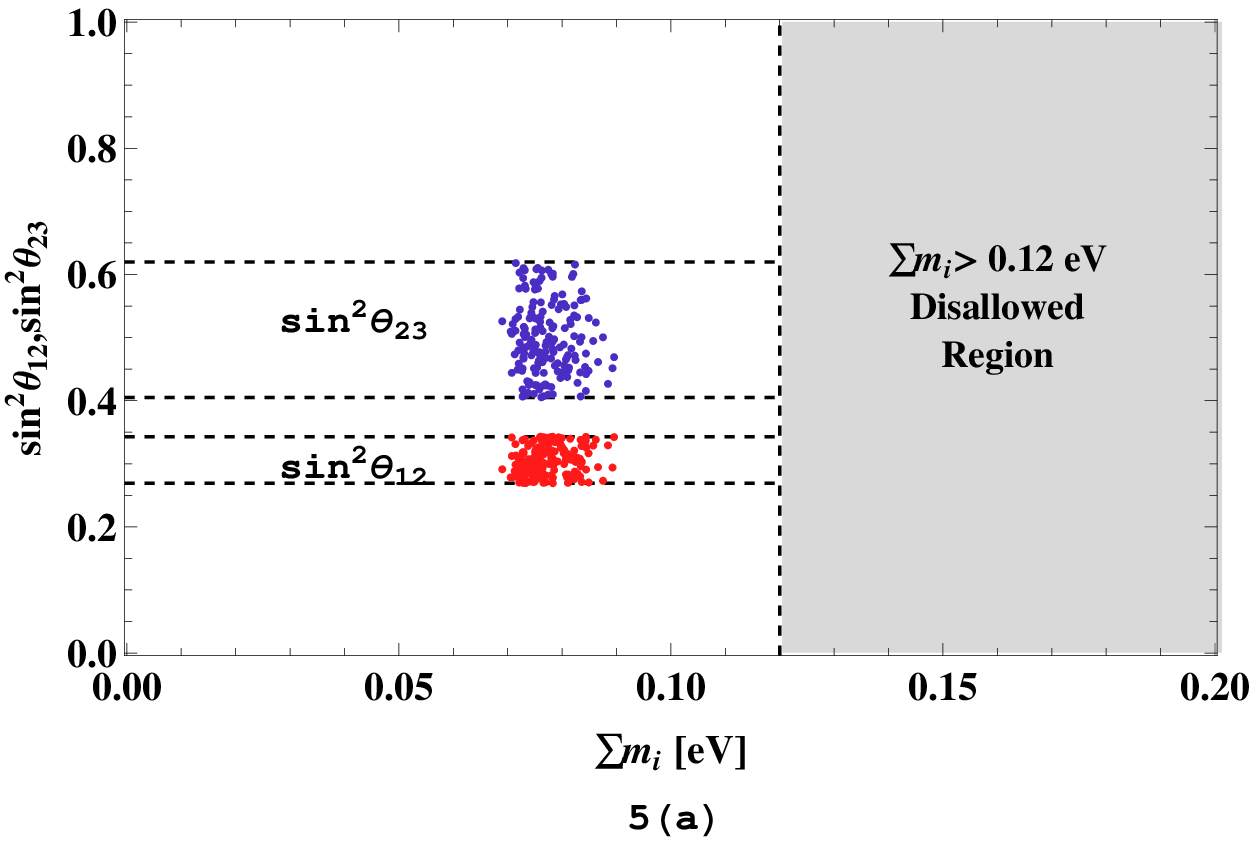} }}%
    \qquad
    {{\includegraphics[width=7cm]{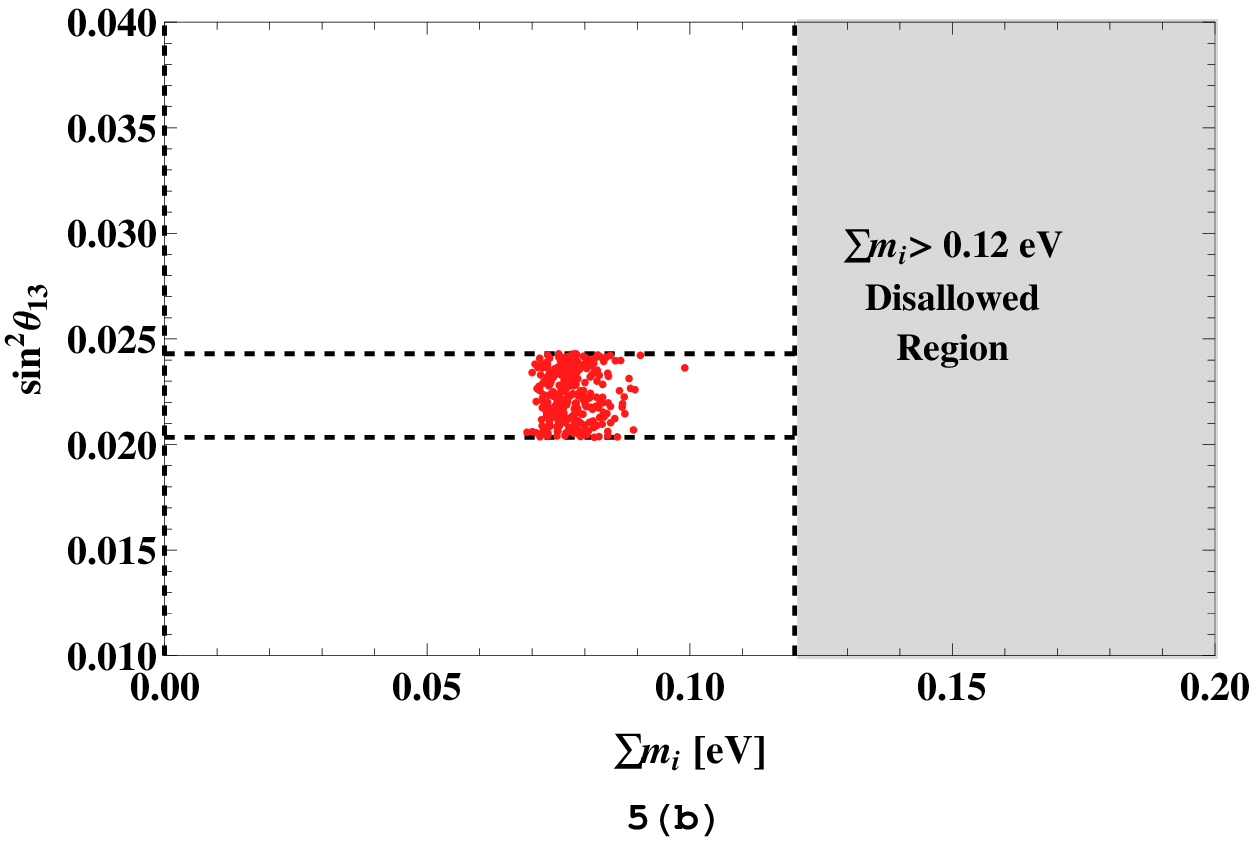}}}
    \caption{The correlation of $\sin^2\theta_{12},\sin^2\theta_{23}$ and $\sin^2\theta_{13}$ with sum of neutrino masses $\Sigma m_{i}$. The horizontal lines are allowed $3\sigma$ ranges of mixing angles \cite{Esteban:2020cvm} and the grey shaded region is disallowed by cosmological bound on sum of neutrino masses \cite{Giusarma:2016phn,Planck:2018vyg}.}%
    \label{fig4}%
\end{figure}

\subsection{Numerical Analysis}

In addition to the parameters in Table \ref{tab2}, we randomly vary $M_\psi$ and $y_{\psi}$ in the ranges [100, 400] GeV and [1, 3.544], respectively. Following the procedure as described in Sec. \ref{sec2.2}, we numerically diagonalize the low energy effective neutrino mass matrix Eqn. (\ref{mmatrix}) to obtain predictions on neutrino masses and mixing angles. For the sake of completeness, we have given some correlation plots depicting the allowed parameter space of the model. In Fig. \ref{fig4}(a) and \ref{fig4}(b) we have shown the correlation plots of $\sin^{2}\theta_{12}$/$\sin^{2}\theta_{23}$ and $\sin^{2}\theta_{13}$ with $\Sigma m_{i}$, respectively. It is evident that the model predicts neutrino mixing angles in their $3\sigma$ ranges consistent with the latest neutrino oscillation data\cite{Esteban:2020cvm}.

Also, information about the $CP$ violation is encoded in $CP$ rephasing invariants $J_{CP}$, $I_1$ and $I_2$. The Jarlskog $CP$ invariant $J_{CP}$ is given by \cite{Krastev:1988yu,Jarlskog:1985ht} 

\begin{equation}
    \text{$J_{CP}$} = \text{Im}[U_{e1}U_{\mu 2}U_{e2}^{*}U_{\mu 1}^{*}],
\end{equation}
while the other two $CP$ invariants $I_{1}, I_{2}$ related to Majorana phases can be written as

\begin{align}
 I_{1} = \text{Im}|U_{e1}^{*}U_{e2}| ,\quad \quad I_{2} = \text{Im}|U_{e1}^{*}U_{e3}|.
\end{align}

\begin{figure} [t]%
    \centering
    {{\includegraphics[width=7.5cm]{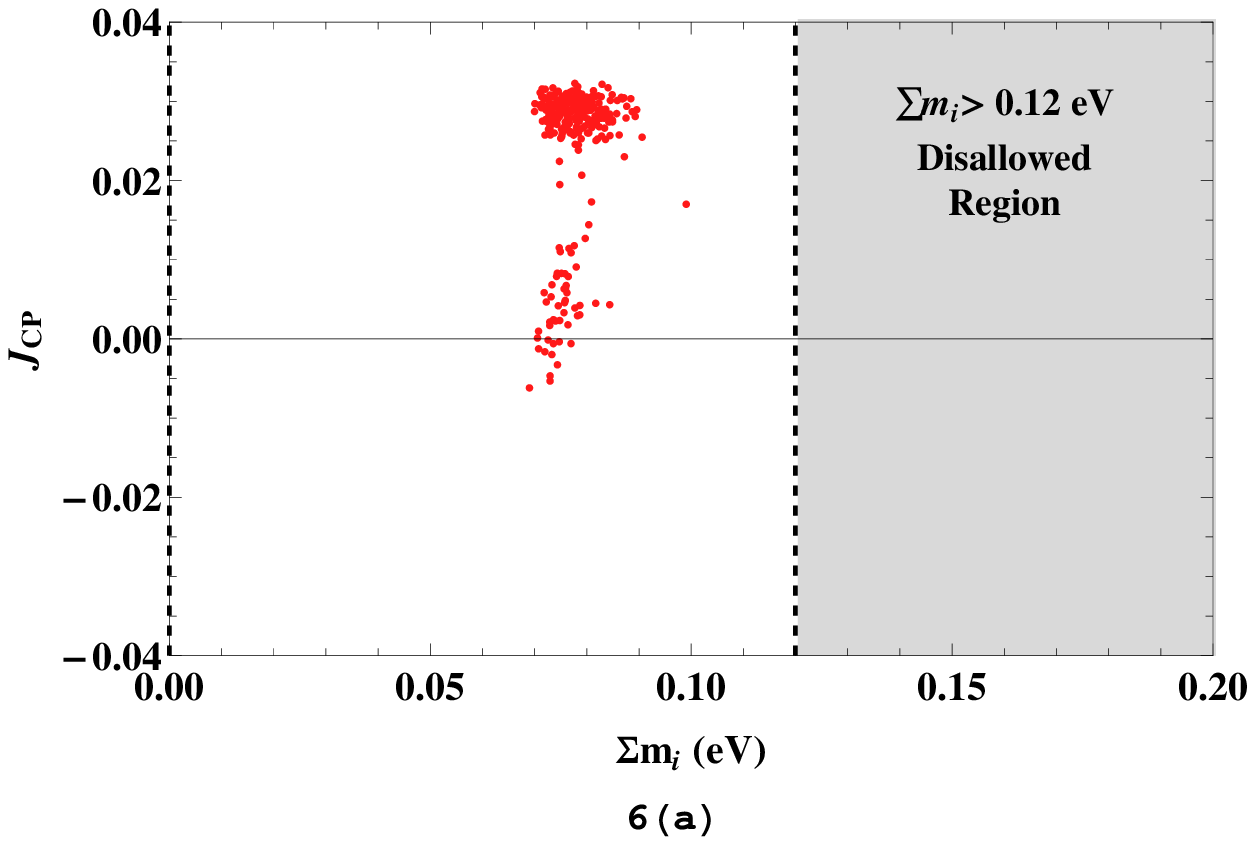} }}%
    \qquad
    {{\includegraphics[width=7.5cm]{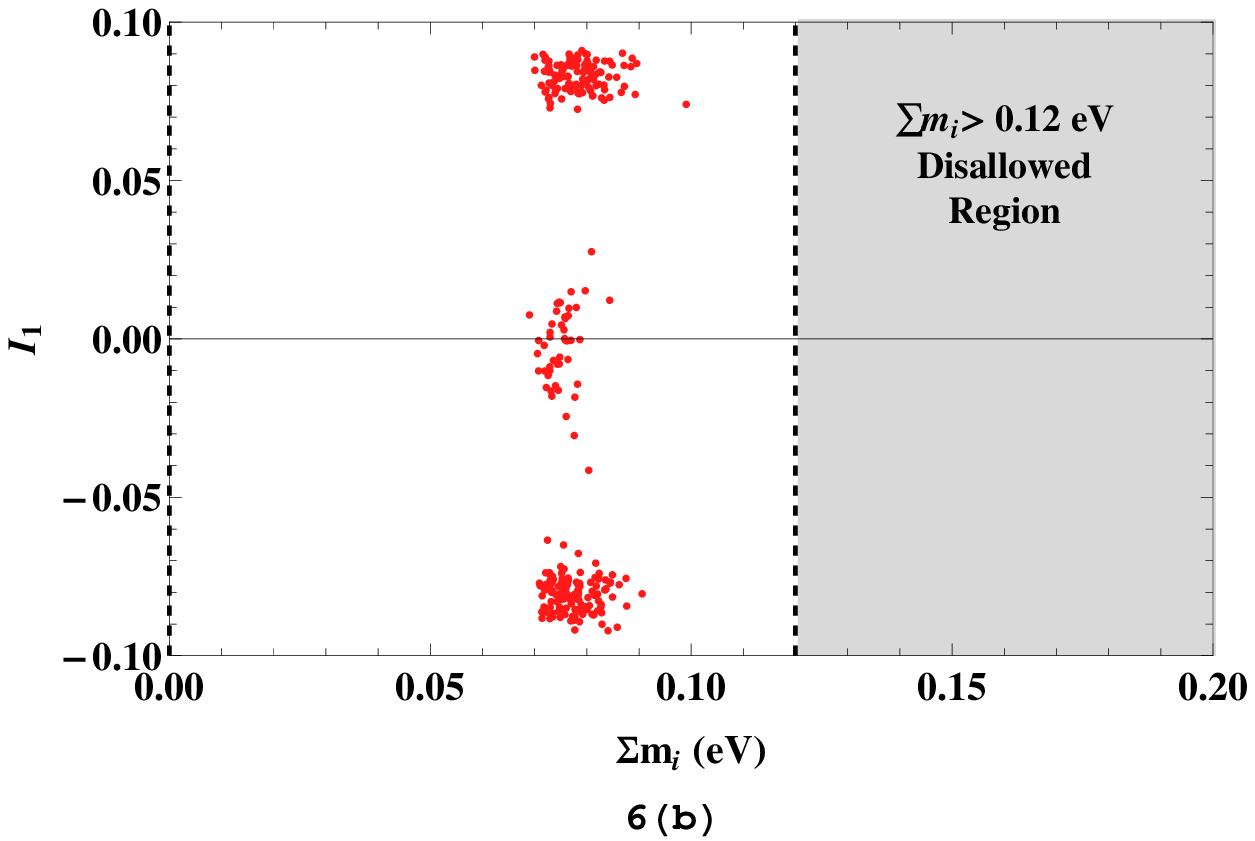}}}
    {{\includegraphics[width=7.5cm]{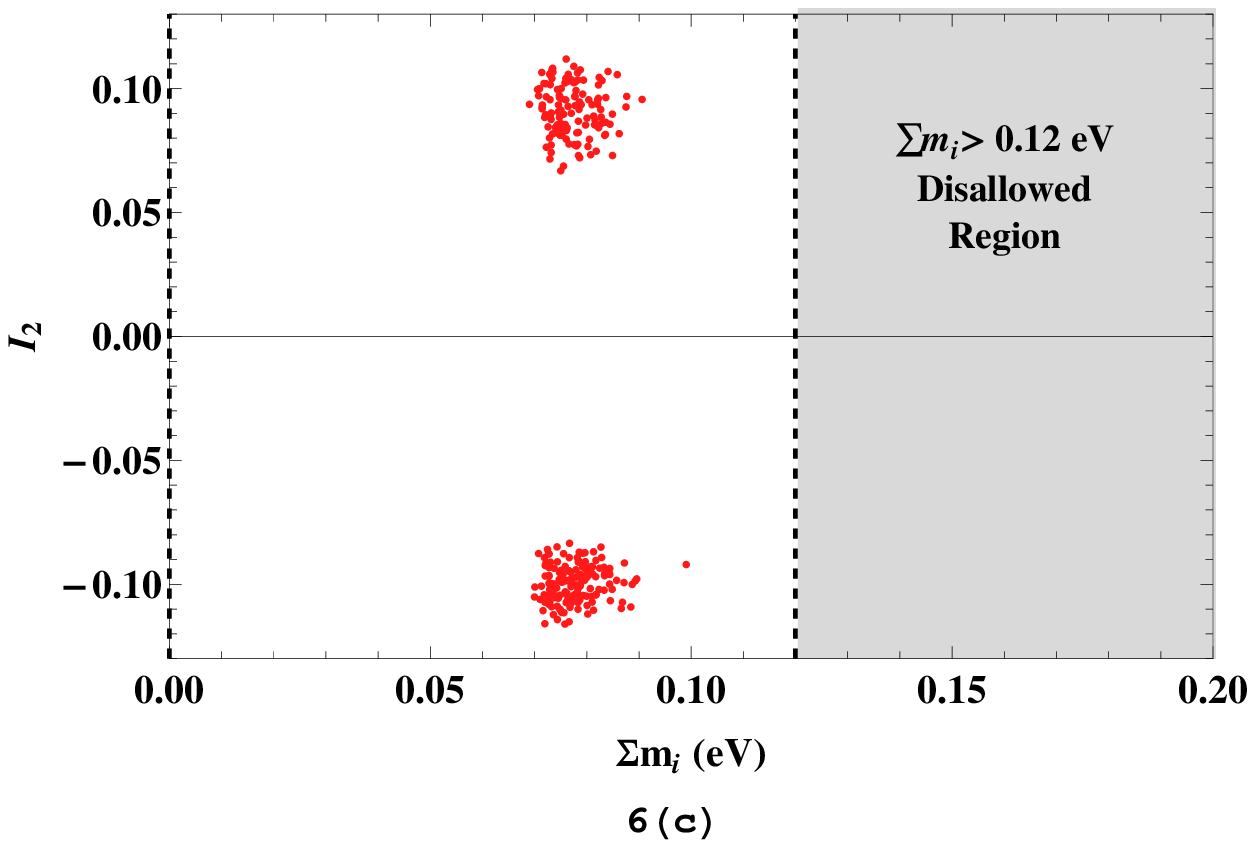} }}%
    \qquad

    \caption{The correlations of $CP$ rephasing invariants $J_{CP},I_{1}$ and $I_{2}$ with sum of neutrino masses $\Sigma m_{i}$. The grey shaded region is disallowed by the cosmological bound on sum of neutrino masses \cite{Giusarma:2016phn,Planck:2018vyg}.}%
    \label{fig6}%
\end{figure}

Fig. \ref{fig6}(a) shows the variation of Jarlskog $CP$ invariant with sum of active neutrino masses $\Sigma m_{i}$ and Figs. \ref{fig6}(b) and \ref{fig6}(c) show the correlation of $I_1$ and $I_2$ with sum of active neutrino masses $\Sigma m_{i}$. The model predicts both $CP$ conserving and violating solutions.  

Furthermore, there is a longstanding question in particle physics about the exact nature of neutrinos. Neutrinoless double beta decay ($0\nu\beta\beta$) process can shed light on whether neutrino is Dirac or Majorana particle. Other than the phase space factor, amplitude of this process is proportional to (1,1) element of the neutrino mass matrix (Eqn. (\ref{mmatrix})). We, also, calculate the model prediction for this process. The effective Majorana mass appearing in the $0\nu\beta\beta$ decay can be written as
\begin{equation}
    m_{ee}  \equiv  M_{11}^{\nu}  = \left|\sum_{i=1}^{3} U_{ei}^{2} m_{i} \right|.
\end{equation}

\begin{figure} [t]%
\centering
    {{\includegraphics[width=7.5cm]{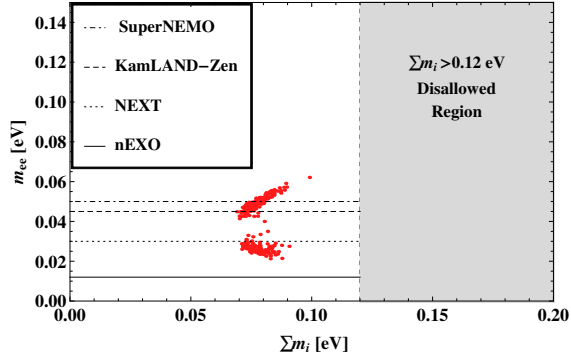}}}
    \caption{The correlation of $m_{ee}$ with sum of active neutrino masses $\Sigma m_{i}$. The sensitivity reach of various $0\nu\beta\beta$ decay experiments are shown as the horizontal lines. The grey shaded region is disallowed by the cosmological bound on sum of neutrino masses \cite{Giusarma:2016phn,Planck:2018vyg}.}%
    \label{fig5}%
\end{figure}

 Fig \ref{fig5} shows the correlation  of effective Majorana mass $m_{ee}$ with $\Sigma m_{i}$. Several $0\nu\beta\beta$ decay experiments with high sensitivities such as nEXO \cite{Licciardi:2017oqg}, NEXT \cite{NEXT:2009vsd,NEXT:2013wsz}, KamLAND-Zen \cite{KamLAND-Zen:2016pfg} and SuperNEMO \cite{Barabash:2011row} have bright prospect for its observation. The sensitivities of these experiments are, also, shown in Fig. \ref{fig5} which have imperative implication for the model. For example, the non-observation of $0\nu\beta\beta$ decay down to the sensitivity of 0.01 eV will refute the model.

\textbf{Muon ($g-2$):}

\begin{figure}[t]
    \centering
    \includegraphics[width=7.5cm]{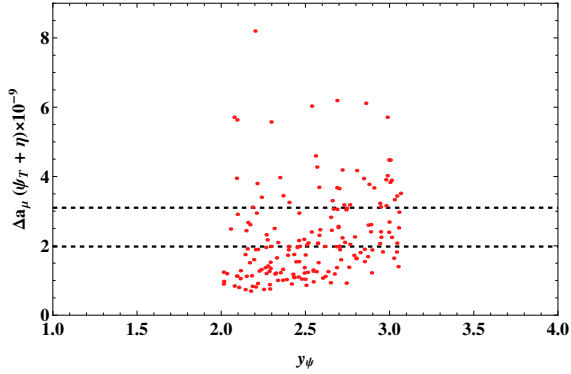}
    \caption{The prediction for muon anomalous magnetic moment, considering mixing of $\psi_T$ and $\eta$,  in $y_\psi$-$\Delta a_\mu$ plane. The horizontal lines show the experimental range of $\Delta a_{\mu}=a_\mu^{exp}-a_\mu^{th}=(2.51\pm 0.59)\times 10^{-9}$\cite{Muong-2:2021ojo}.}
    \label{fig8}
\end{figure}

It is well known that, in gauged $U(1)_{L_{\mu-\tau}}$ models, the $vev$ of scalar which breaks this symmetry is chosen such that $M_{Z_{\mu\tau}}$ lies within muon neutrino trident (MNT) upper bound of 300 MeV to satisfy muon ($g-2$). It is to be noted that though neutrino phenomenology can be satisfied for higher $M_{Z_{\mu\tau}}$ as discussed in Sec \ref{sec2.2} but it does not provide a solution to $\Delta a_{\mu}$. The contribution to $a_{\mu}$ for this scenario is provided by coupling of $\psi_{T}$ with $\eta$ through Eqn. (\ref{au2}). In Fig. \ref{fig8} we have shown the region of parameter space contributing to anomalous magnetic moment of muon $\Delta a_\mu$. The horizontal lines depict the experimental allowed range of $\Delta a_\mu$.

\begin{table}[t]
   
      \centering
        \begin{tabular}{ll}
             \hline
             \hline
       Parameters  &  Value \\
       \hline
       
        ($M_{ee}$,  $M_{\mu\tau}$) & (62.2, 73.2) GeV \\ 

          ($x, y$) &  (58.9, 27.1) GeV \\ 
         $ \delta$ & 282.3 \\ 
             $y_{\eta k}$  & (0.008, 0.008, 0.008)  \\
                ($M_{\eta_{R}^{0}}, M_{\eta_{I}^{0}}$) & (278..21, 278.22) GeV \\ 
                  $v_{S}$ & 870 GeV \\
                     $ N_{k}$ & (6.53, 1.41, 5) $\times 10^{6}$ GeV  \\ 
                       \hline
        \end{tabular}
   
    \caption{The values of input parameters used to obtain benchmark points of numerical analysis for the extended model ($k=e,\mu,\tau$).}
    \label{tab4}
\end{table}

\textbf{Benchmark Point: } For ready reference, we provide benchmark point showing the viability of the model to predict neutrino mixing angles and mass squared differences within experimental range and, also, to explain muon ($g-2$) using fermion triplet $\psi_{T}$. For input parameters as listed in Table \ref{tab4}, neutrino mixing angles and mass squared differences as predicted by the model are 

 \begin{equation*}
    \sin^{2} \theta_{13} = 0.023 ,\quad  \sin^{2} \theta_{23} = 0.49  ,\quad \sin^{2} \theta_{12} = 0.32
    \end{equation*}
    \begin{equation*}
  \Delta m^{2}_{31} = 0.0024 eV^{2} , \quad \Delta m^{2}_{21} = 0.000073 eV^{2},
    \end{equation*}
and the contribution to muon anomalous magnetic moment $\Delta a_{\mu}(\psi_{T} + \eta )=3.08\times 10^{-9} $ for $M_{\psi} = 329$ GeV, $M_{\eta} = 126$ GeV and $y_{\psi} = 2.74$.

 \section{Conclusions}

In conclusion, muon ($g-2$) anomaly, non-zero neutrino mass, nature of neutrinos lacks an explanation within the SM. In this work, we revisit the gauged $U(1)_{L_{\mu}-L_{\tau}}$ extension of the SM with in the framework of scotogenic model. The particle content of the model is extended by three right handed neutrinos, one inert scalar doublet and a SM gauge singlet scalar to implement conventional scotogenesis. $U(1)_{L_{\mu}-L_{\tau}}$ symmetry is broken by the $vev$ $v_S$ resulting in massive gauge boson $Z_{\mu\tau}$. We have shown that the model predicts the neutrino oscillation parameters with in their experimental range and simultaneously explains the muon ($g-2$). The mass of the gauge boson $M_{Z_{\mu\tau}}=g_{\mu\tau}v_S$ is constrained by the neutrino phenomenology (through $M_R$). In fact, we obtain two distinct regions (around 150 MeV and 500 MeV) in $M_{Z_{\mu\tau}}-g_{\mu\tau}$ plane which can explain the neutrino phenomenology. Region I is consistent with the MNT bound, however, region II violates it for mass range $M_{Z_{\mu\tau}}>300$ MeV. The explanation of muon $g-2$ based on gauged scotogenic model assumes small mixing between $Z_{\mu\tau}$ and SM gauge boson $Z$. In Sec. \ref{sec3}, we propose an extension of the gauged scotogenic model with VLL triplet wherein muon ($g-2$) can be explained independent of above mixing pattern and $U(1)_{L_\mu-L_\tau}$ gauge boson mass. In this case we have shown that, in light of the neutrino oscillation data, muon ($g-2$) can be explained via mixing between VLL triplet $\psi_T$ and inert scalar doublet $\eta$ in the mass range (100-400) GeV. We have, also, given the benchmark points for both the scenarios emphasising the viability of the model. In addition, the implication of the model for $0\nu\beta\beta$ decay has, also, been studied. The non-observation of $0\nu\beta\beta$ decay down to the sensitivity of 0.01 eV shall refute the model. The model, in general, predicts both $CP$ conserving and violating solutions.     
\vspace{1 cm}

\noindent \textbf{\Large{Acknowledgments}}\\
S. Arora acknowledges the financial support provided by the Central University of Himachal Pradesh. M. K. acknowledges the financial support provided by Department of Science and Technology, Government of India vide Grant No. DST/INSPIRE Fellowship/2018/IF180327. The authors, also, acknowledge Department of Physics and Astronomical Science for providing necessary facility to carry out this work. B. C. Chauhan is thankful to the Inter University Centre for Astronomy and Astrophysics (IUCAA) for providing necessary facilities during the completion of this work.


\end{document}